# Assessing the Utility of Structure in Amorphous Materials

Dan Wei, Jie Yang, Min-Qiang Jiang, Lan-Hong Dai and Yun-Jiang Wang

*State Key Laboratory of Nonlinear Mechanics, Institute of Mechanics, Chinese Academy of Sciences, Beijing 100190, P. R. China; School of Engineering Science, University of Chinese Academy of Sciences, Beijing 101408, P. R. China*

Jeppe C. Dyre

*Glass and Time, IMFUFA, Department of Science and Environment, Roskilde University, P. O. Box 260, DK-4000 Roskilde, Denmark*

Ian Douglass and Peter Harrowell

*School of Chemistry, University of Sydney, Sydney N.S.W. 2006, Australia*

Abstract

This paper presents a set of general strategies for the analysis of structure in amorphous materials and a general approach to assessing the utility of a selected structural description. Measures of structural diversity and utility are defined and applied to two model glass forming binary atomic alloys. In addition, a new measure of incipient crystal-like organization is introduced, suitable for cases where the stable crystal is a compound structure.

## 1. Introduction

The explanation of material properties and behaviour in terms of the microscopic structure constitutes the *modus operandi* of the physical sciences – chemistry, physics and materials science. It seems a natural expectation, therefore, that a science of amorphous materials should, eventually, be built on analogous structural explanations. While a considerable body of literature [1,2] records the effort to advance just this program, success has proven elusive. The first problem is to identify exactly what are we referring to as 'structure' in an amorphous material. With their periodic repetition of a single unit cell, crystal structures require only a small amount of information to specify the total structure. This is not the case in amorphous solids, where any useful measure of structure (where 'useful' refers to a measure that does not involve a complete specification of every particle position) must be seriously incomplete. It follows that each specific structural measure will unavoidably represent a choice regarding what information is retained and what is discarded. Some choices must be more useful than others. In this paper we consider how one might assess the utility a structural measure of an amorphous material.

For our purposes, the 'structure of an amorphous material' shall be taken to mean the frequencies of some local classification of particle configurations. This is the definition of amorphous structure that is almost universally used in the literature [1,2]. Access to this information is largely via models. Initially, the models were real: the bubble rafts of Bragg and Nye [3] and Bernal's randomly close packed ball bearings [4]. With the advent of computer simulations, these analog models were replaced by digital ones. Some important early examples of the application of Voronoi polyhedra to the analysis of simulated liquid structures are the papers by Rahman [5] and Tanemura et al [6].

Our definition of structure requires that the researcher must make a choice of classification scheme. This choice is integral to the requirement that structure be intelligible. The choice of the local classification criteria is unrestricted. Along with Voronoi polyhedra, examples of local classification include common neighbours [7], coordination geometries of nearest neighbours (roughly, the dual of the Voronoi representation) [8], clusters (e.g. poly-tetrahedrality) [9], local ring lengths [10] and degree of local centro-symmetry [11]. The question we address in this paper is how useful is a given structural representation given the arbitrary choice that is made regarding how local structure is classified? Many of the papers reporting on the structure of an amorphous material assume that structure, however defined, is useful by default. The object of this study is to establish how this assumption can be put to a meaningful test.

'Useful' is probably even more treacherous a concept to define than 'structure'. To what uses do we put structural information? In this paper we shall consider the following three. i) *Structure as information compression*. The information required to determine the positions of all of the atoms in a crystal is typically small (i.e. the unit cell structure and lattice parameters) and independent of the size of the sample. This dramatic data compression provided by the structure allows for the structure of crystalline materials to be easily stored, recovered and used. ii) *Structure as a casual explanation of a physical property*. The notion of energetically favoured local structures is a common starting point for rationalising liquid structure. Malins et al [12] have used this energetic criterion as means of identifying coordination polyhedral of interest in resolving amorphous structure. If the stability of an amorphous material could be closely correlated to the presence of specific local structures, these structures could, in turn, provide an explanation of a range of material properties arising from configurational stability. iii) *Structure as a measure of proximity to an ordered phase*. One of the most cited papers on amorphous structure is a short note by Frank [13] in 1952 in which he suggested that local icosahedral coordination shells might stabilize a pure liquid metal sufficiently to allow it to be supercooled. As developed by Mackay [14], Hoare and Pal [15], Kivelson et al [16], the idea has evolved into a proposition that disorder may be underpinned by a form of geometrically frustrated order.

There is a fourth common usage of structure –*Structure as the rules by which the whole is assembled from its parts*. While this can be regarded as an example of data compression (i.e. the first point in our list), assembling a structure is a quite specific process and one of the more generic characteristics of a structure (as something that has been assembled). We shall not consider this use since the standard definition of amorphous structure in terms of the distribution of local environments discards, by construction, the information about the correlations between local structures and, hence, how these local environments are assembled. There have been studies that have sought to extend the structural characterisation to include the spatial arrangement of coordination polyhedral [17] but it remains unclear whether the accumulation of the additional information needed for these extended descriptions is rewarded by an increased insight into the amorphous material. In looking for a more compact description of extended structure, some workers have considered approximating the extended arrangement of local coordination polyhedra as a disordered (plastic) crystal lattice [18,19]. Whether such approximants are even stable with respect to the non-periodic reality has been questioned [20].

## 2. Models and Algorithms

In this paper we shall use two well studied model glass forming liquids, both based on binary atomic mixtures. One is a model of CuZr using a many body potential of the Embedded Atom type due to Mendelev et al [21]. The equilibrium crystal phase of the model CuZr [22], is the B2 structure (i.e. a body centred cubic structure with the two species occupying alternating sites similar to that found in CsCl). While the B2 crystal has been observed to grow [23], the simulated CuZr has proven highly resistant to nucleation during extended simulations. The second model is a mixture of two Lennard-Jones particles introduced by Kob and Andersen (KA) [24]. The interparticle potential has the following functional form

$$\phi_{ab}(r) = 4\varepsilon_{ab}\left[\left(\frac{\sigma_{ab}}{r}\right)^{12} - \left(\frac{\sigma_{ab}}{r}\right)^{6}\right] \quad (1)$$

with potential parameters $\sigma_{AA}= 1.0$, $\sigma_{AB} = 0.8$ and $\sigma_{BB} = 0.88$, and $\varepsilon_{AA} = 1.0$, $\varepsilon_{AB} = 1.5$ and $\varepsilon_{BB} = 0.5$. For the KA model we shall use temperature units of $\varepsilon_{AA}/k_B$, energy units of $\varepsilon_{AA}$ and length units of $\sigma_{AA}$. The model has mainly been studied at the $A_{80}B_{20}$ composition (the same composition associated with optimal glass formation in NiP [25] on which the KA model was based). The equimolar mixture crystallizes readily in the same B2 crystal structure as found in the CuZr model [26,27]. At $A_{80}B_{20}$ crystallization is much slower and is driven by crystallization of the face centred crystal of pure A [27]. The structure of the supercooled liquid in CuZr [28] and the KA mixture [29] have both been studied extensively.

The amorphous states studied in this paper were generated by a continuous cooling of a liquid initially equilibrated above its melting point down to T = 0 under the constraint of constant pressure. The resulting energy minimum can be characterised by a fictive temperature equal to the temperature at which, on cooling, a property such as the volume deviated from the equilibrium value as a consequence of structural arrest. In Fig. 1 we plot the dependence of the volume V for the CuZr and $A_{80}B_{20}$ alloys as a function of temperature during cooling and indicate the temperature at which the deviation from equilibration occurs in each case and, hence, the fictive temperature of the respective T = 0 glasses.

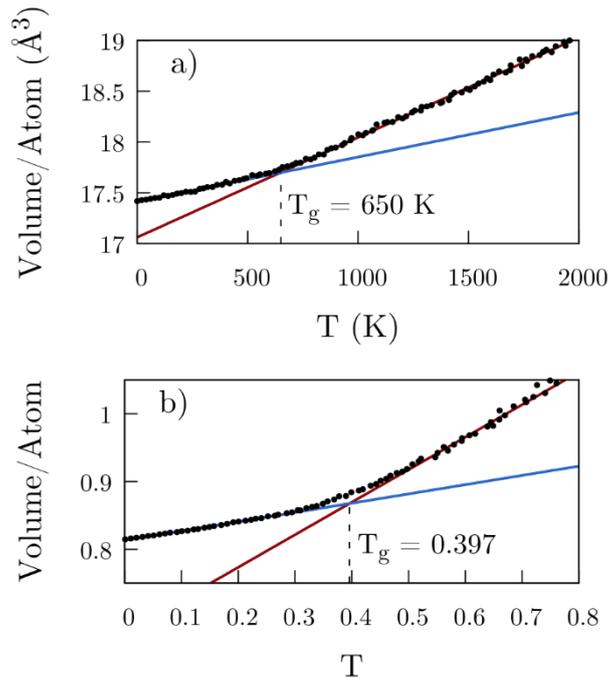

**Figure 1.** The glass transition temperatures $T_g$ for a) CuZr and b) the KA mixture at $A_{80}B_{20}$ for the cooling rates $10^{10}$ K/s and 1.3 x $10^{-5}$, respectively, used to generate the T=0 configurations whose structures are reported in this paper. $T_g$ is defined as the temperature at which dV/dT undergoes an abrupt change. This $T_g$ is the fictive temperature of the T=0 configuration. The units in b) are the Lennard-Jones reduced units.

The local structure of the T=0 amorphous states has been characterised using Voronoi polyhedral. We have used a standard 4 digit descriptor for the polyhedral $(n_3, n_4, n_5, n_6)$ where $n_i$ is the number of facets with i edges. Note that this Voronoi analysis is purely topological and does not differentiate the two species in the local coordination shell. The Voronoi analysis suffers from a problem common to many forms of structural classification defined in terms of neighbour separations. The identification of a neighbour is all-or-nothing depending on some threshold distance – explicit or implicit – resulting in substantial fluctuations in the topological signature due to the fluctuations of separations close to this threshold. In the case of Voronoi analysis, large separations correspond to small faces, a problem that has been discussed previously [30]. An example of this issue is the Voronoi structure of the FCC crystal at a non-zero temperature. The Voronoi polyhedron around an atom in a perfect FCC lattice is (0,12,0,0) but this polyhedron is not observed at finite T – instead the dominant polyhedron is (0,6,0,8) – one characteristic of the BCC lattice – simply as a consequence of vibrational motion (i.e no defects are required to see this structural broadening). An alternative approach has been introduced [31] using Minkowski tensors that weight neighbour contributions based on their separation from the central particle that promises to reduce these fluctuations. While we shall not explore these more sophisticated measures – in this paper we seek to frame general questions about structure - readers are encouraged to view the statistics of any local structural description as a combination of real local variability and the noise imparted by the details of the chosen metric.

## 3. Results

### 3.1 Structure as Data Compression: On Quantifying Diversity

The statistical structure of a glass, as we have defined it here, takes the form of the fraction $p_i$ of particles in the local polyhedral labelled $i$. In Fig.2 we present these fractions for CuZr and the KA mixture with composition $A_{80}B_{20}$. Also presented, for comparison, is the analogous structure of the B2 crystal structure as formed from the quenched $A_{50}B_{50}$ KA mixture. This crystal is the equilibrium crystal phase for the model CuZr alloy as well. In both amorphous alloys we find a broad distribution of local structures, with no one structure exceeding 7% in frequency. This flat distribution of Voronoi structures is a common place observation for amorphous alloys [1,2]. In the absence of any outstanding structures, we suggest that the most striking feature of distributions like those shown in Fig. 2 is exactly their multiplicity. Indeed, the most straightforward and general structural differentiation between a crystal and a glass is not the presence of any specific structure but the variety of local structures in the latter as compared to the former. Quasicrystals, for example, lack periodicity but are still ordered by virtue of consisting of only a small number of local structures [32].

We can quantify the multiplicity of the distributions in Fig. 2 as follows. The Shannon information S [33] associated with a particular classification (e.g. Voronoi, common neighbour, etc) is given by

$$S = -\sum_i p_i \ln p_i \tag{2}$$

The diversity D of structures – i.e. the effective number of distinct structures present – is related to the information S through the relation

$$D = \exp(S) \tag{3}$$

This definition of diversity is used in studies of animal and plant populations [34]. The values of S and D are included on each graph in Fig.2. For the perfect B2 crystal there is a single local structure, so S = 0 and D = 1. The difference between this ideal and the values reported for the crystal as formed in Fig. 2c is that defects have been captured in the crystal as it was crystallized during the simulation. With values of D = 95 and 126, for amorphous CuZr and $A_{80}B_{20}$ respectively, it is clear that the Voronoi classification of the amorphous alloys leaves us with a very large diversity of local structures. To appreciate just how large these amorphous D's are, it is helpful to consider the range of diversity in crystals. A survey of over 16000 intermetallic crystal structures [35] reports that over 92% had 4 or less distinct coordination geometries (i.e. D ≤ 4). Note that the unit cells can be much larger than the number of distinct coordination sites as the same site might appear in different orientations. Based on this statistic, we might tentatively suggest that crystals correspond to a structure class characterised by a low (i.e. D <10) structural diversity. An analogous observation was been enshrined by Pauling in his 'rule of parsimony' [36].

The diversity, as defined here, is distantly related to the configurational entropy. Both quantities reflect our efforts to enumerate configurations in terms some sort of imposed resolution. Where diversity depends on a researcher's choice of structural measure, configurational entropy [37] depends on the researcher's definition of a reference configuration (e.g. local potential energy minima, distinct free energy minima, etc). Both measures provide a useful insight as to how the multiplicity of configurations decrease with cooling. The diversity expresses that multiplicity explicitly in terms of the types of structure selected.

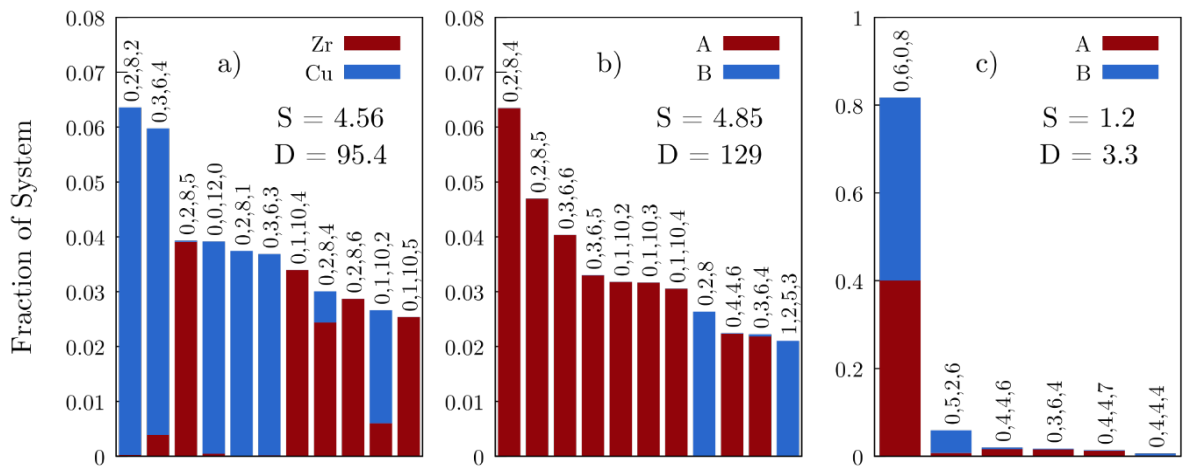

**Figure 2.** Fraction of atoms in environments classified by various Voronoi polyhedral for a) CuZr, b) $A_{80}B_{20}$ and c) the B2 crystal as crystallised from the equimolar $A_{50}B_{50}$ KA liquid. The colors red and blue represent the identity of the central atom as indicated.

The diversity D is introduced here as a useful tool for quantifying structures that might be more usefully characterised by their multiplicity rather than by the frequencies of a few individual structures. As an example of how we might employ the diversity D, let's consider how the diversity of structures changes as we increasingly restrict our consideration to those structures associated with some extreme of a property. We shall consider the degree of constraint experienced by particles as measured by $<\Delta r^2>/k_B T$ where $<\Delta r^2>$ is calculated by averaging over trajectories at a T well below the glass transition. We are generally interested in the structures corresponding to high constraint. To this end we shall pick a threshold value $<\Delta r^2>^*/k_B T$ and then determine the histogram of structural frequencies for the subpopulation of particles for which

$<\Delta r^2>/k_B T \; < \; <\Delta r^2>^*/k_B T$. In Fig. 3 we plot the variation of D with the choice of threshold value $E^*$ and $<\Delta r^2>^*/k_B T$ for the two mixtures. The null hypothesis is that the constraint is independent of structure and so randomly distributed across all local structures. If this were so then the change in the threshold would not change the diversity of structures at all, just the size of the subpopulation being sampled.

The effect of sample size on D is an important practical problem that we need to address. It is obvious that as our sample size becomes small enough that the less frequent structures are not being sampled at all, then the diversity will drop until, in the limit of a sample size of one, the diversity must, by definition, vanish and D =1. To assess the effect of sample size, we have included in Fig. 3 curves associated with the null hypothesis for which we generated subpopulations of the same size as the one generated by applying the threshold but with particles randomly selected. The uncertainty in D due to finite sampling can be measured as the difference |$D_{total}$ -$D_{null}$| where $D_{total}$ is the diversity calculated over the total system while $D_{null}$ is the diversity of a randomly selected group of particles from a subpopulation the same size as that generated by applying the threshold. We have indicated on the plots in Fig. 3 where the sampling error becomes too large (i.e. >10% of the calculate D) to represent usable data. Within the range of statistically significant data we find a substantial decrease in the diversity, with D decreasing from 96 to 56 in CuZr and from 129 to 37 in $A_{80}B_{20}$. This means that constraint does exert a significant degree of selectivity on structure, while still leaving a large number of contributing structures.

One contributing factor to this reduction in diversity is that the constraint favours one species over the other. In Fig. 4, we plot the change composition of the sub-population as a function of $k_B T / <\Delta r^2>^*$. We find that the larger particles are more strongly represented in the high constraint particles than the smaller ones. In the CuZr mixture, the fraction of Zr centred structures for which $0.65 \leq k_B T / <\Delta r^2>^*$ is 0.77 (as compared to 0.5 for the total system). In the $A_{80}B_{20}$ mixture, the fraction of the larger A particles for which $55 \leq k_B T / <\Delta r^2>^*$ is 0.98 (again, a significant increase over the total value of 0.8). The loss of diversity associated with the complete loss of small particles is given by the difference $D_{total}$-$D_{large}$ (where $D_{large}$ is the diversity of the large particles measured across all values of constraint) which, as plotted in Fig. 3, is 19 and 40, for CuZr and $A_{80}B_{20}$, respectively. Assuming a linear interpolation based on the measured change in composition, we estimate the fraction of the maximum decrease in diversity (i.e. for the maximum values of $k_B T / <\Delta r^2>^*$ for which statistical significance is retained) that is due to the change in composition: 26% and 51% for the CuZr and $A_{80}B_{20}$ mixtures, respectively. The remaining loss of diversity in each case is due to the explicit structural selectivity associated with the application of the threshold.

This observation – that structural diversity decreases when we focus on increasingly extreme values of some property – is a generic one in amorphous materials and is often cited as evidence that the favoured structure must have 'caused' the property extreme. As we shall discuss in the following Section, this extraction of causal connection from simple correlations is a non-trivial task.

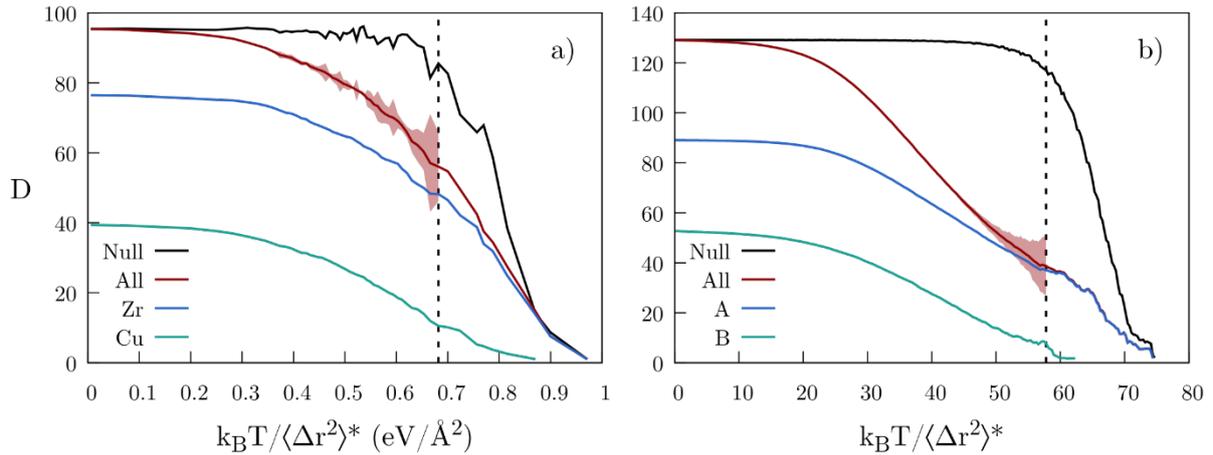

**Figure 3**. Plot of the diversity D for a) CuZr and b) $A_{80}B_{20}$ glass as a function of $k_BT/\langle\Delta r^2\rangle^*$, where $\langle\Delta r^2\rangle^*$ is the upper bound on the mean squared displacement. For each glass-former, values of D are calculated for the individual chemical components as well as for the two components combined as indicated. The null hypothesis (see text) is also plotted. By construction $D_{null}$ only decreases due to finite size sampling errors. Error bars associated with the finite sampling error are calculated as the difference $|D_{total} - D_{null}|$ as explained in the text. The dashed vertical lines in each plot mark the point at which the error exceeds 10% of the calculated D.

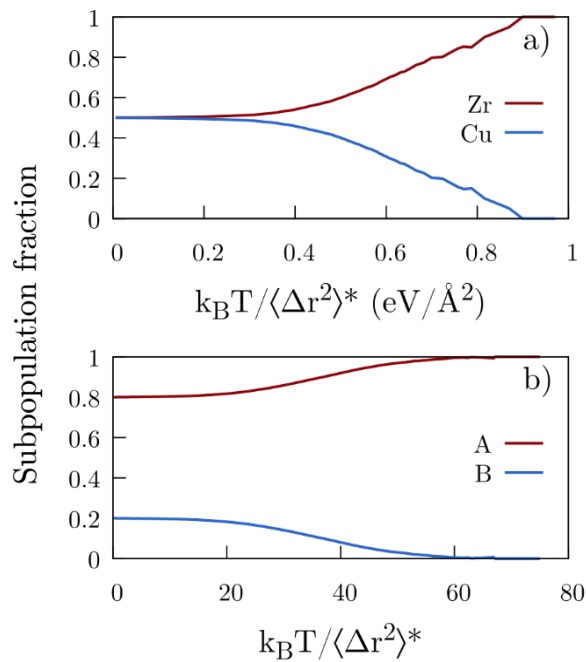

**Figure 4.** Relative contribution of large (i.e. Zr or A) and small (i.e. Cu or B) atoms to the subpopulations depicted in Fig. 3 plotted against $\langle\Delta r^2\rangle^*/k_BT$ for the a) CuZr and b) $A_{80}B_{20}$ mixtures.

## 3.2 Structure and The Causal Explanation of Material Properties

The statement that correlations do not imply causation is a basic tenet of statistics [38]. Physical sciences, in contrast, quite routinely see correlations, coupled with some physical insight, employed to establish explanations of material behaviour. The question of how this apparent difference is bridged is the basis of a body of literature [39] that starts with the 1921 paper by Wright [40]. In this Section, we shall consider how well the expectation that structure can explain properties is met in the case of amorphous materials. Take the relationship between energy and stability. The global groundstate of most many body systems is crystalline. This observation suggests that certain local structures are lower in energy than others and so expected to more frequent (i.e. favoured) on cooling. In general, we propose that the utility of a local structure classification in providing a casual explanation of a property of the material must depend on how well the geometrical classification rules correlate with the property in question.

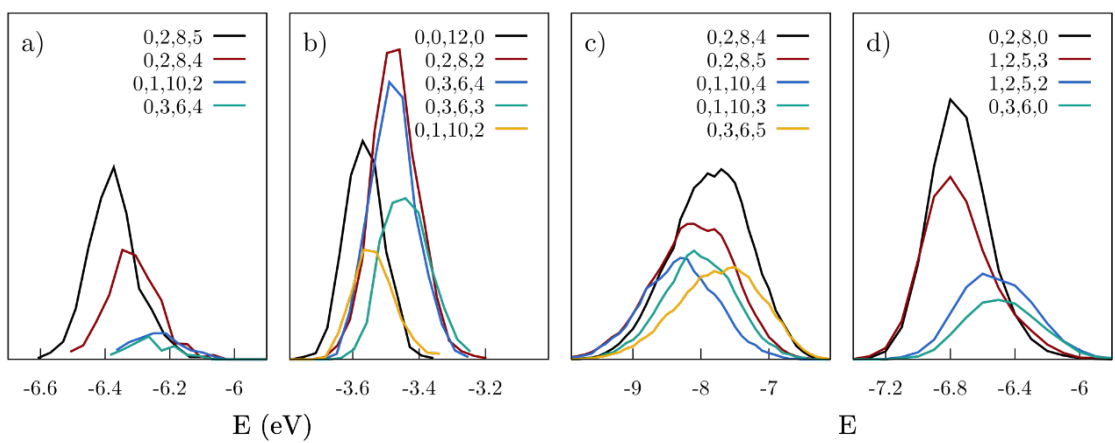

**Figure 5.** The distribution of potential energy for a variety of local structures (as indicated) for the CuZr mixture, with a) Zr-centred polyhedra and b) Cu-centred polyhedra, and for the $A_{80}B_{20}$ mixture with c) A-centred polyhedra and d) the B-centred polyhedra.

In Fig. 5 the distribution of potential energy is plotted for the most frequent local structures of the two model liquids. The distribution clearly separates about the two atomic species. This is, in part, a consequence of the difference in the number of neighbours between large and small particles. In considering the utility of the structure to 'explain' physical properties we can again imagine a null hypothesis in which the structure has no bearing on the energy of an atom. In this case, the energy distributions for the different structure would be identical and, hence, the chosen structural classification was of zero utility in accounting for the distribution of local energies. The alternate limiting case is one characterised by structure-based distributions are narrow and distinct (i.e. with little overlap). The capability of structure to so effectively resolve some physical property would provide strong support for the utility of that structural measure. Utility refers to the degree to which the knowledge of the specific local structure changes our ability to predict the associated property value. Utility of structure, therefore, can be measured by the overlap $Q_{ij}$ of the distribution $p_i(x)$ and $p_j(x)$ of some property *x* for two sub-populations characterised by structures *i* and *j*. We define the overlap $Q_{ij}$ as

$$Q_{ij} = \frac{\int dx p_i(x) p_j(x)}{\sqrt{\int dx p_i^2(x) \int dx p_j^2(x)}} \quad (4)$$

By the Cauchy-Schwarz inequality one has,

$$0 \leq Q_{ij} \leq 1 \quad (5)$$

If $P_i$ is the relative frequency of structure i, then we can define the weighted average overlap Q by

$$Q = \frac{\sum_{i \neq i} P_i P_j Q_{ij}}{\sum_{i \neq j} P_i P_j} \quad (6)$$

where $0 \leq Q \leq 1$. This inequality follows from the fact that Eq. (6) is a so-called convex combination of numbers between zero and unity (Eq. (5)). The utility $U_x$ of a given choice of structural characterization in terms of its capacity to differentiate the property X can then be defined as

$$U_x = 1-Q \quad (7)$$

reflecting the fact that perfect overlap (i.e. Q = 1) would correspond to all $Q_{ij}$ =1, i.e., identical distributions $p_i(x)$, which would correspond to a useless structural resolution while zero overlap would represent an optimal utility with $U_x$ = 1. In the case of the energy, we have calculated values of $U_E$ for the two atomic species separately, i.e. the $Q_{ij}$'s of Eq. 3 are only taken between structures centred around a given type of atom. The value of $U_E$ presented in Table 1 is the average value for the two species. For the utility $U_C$ associated with the degree of particle localization, we have considered distributions of the variable $<\Delta r^2>/k_B T$. We find the utility of the Voronoi polyhedral to account for the distribution of local energy is 0.21 and 0.15 for the CuZr and $A_{80}B_{20}$ alloys, respectively. Similar values are obtained for $U_C$. These results for $U_E$ reflect the considerable overlap of distributions that we see in Fig. 5. The Voronoi classification is clearly insufficient, on its own, to explain the range of local energies in these amorphous materials. The utility $U_E$ of the Voronoi analysis is found to systematically decrease as the concentration of the KA mixture approaches the equimolar value. This is probably a generic result arising from the increasing number of distinct polyhedral found as the equimolar concentration is approached. This trend is not observed for $U_C$.

|  | **CuZr** | **A$_{50}$B$_{50}$*** | **A$_{66}$B$_{33}$** | **A$_{80}$B$_{20}$** |
|---|---|---|---|---|
| **U$_E$** | 0.21 | 0.08 | 0.13 | 0.15 |
| **U$_C$** | 0.10 | 0.06 | 0.16 | 0.10 |
| **f$_x$** | 0.17 | 0.55 | 0.07 | 0.03 |
| **D** | 95 | 426 | 253 | 126 |

* The data for the equimolar KA mixture was obtained from an instantaneous quench rather than a constant finite cooling rate to avoid crystallization.

**Table 1.** Values of the utility U$_E$ and U$_C$, the common structure fraction f$_x$ and the diversity D for CuZr and three concentrations of the KA mixture.

Inspection of the energy distributions in Fig. 5 makes it clear that if we were to look at a subset of particles with sufficiently low energy we would find them dominated by a small number of specific structures. In Fig. 6 we plot the change in the fraction of the smaller B particle structures in the A$_{80}$B$_{20}$ mixture when considering different sub-populations defined by an energy threshold value E$^*$. We find that the fraction of some structures – (0,2,8,0) and (1,2,5,3) - increase as we focus on the lowest energies while others – (0,6,3,0), (1,5,2,2) and (1,3,3,3) – are selected against and decrease. This is observation is consistent with the notion of favoured local structures. The (0,2,8,0) polyhedra, previously identified as a favoured structure in the KA mixture [41,42], is the bicapped square anti-prism, the basis of the metastable Al$_2$Cu crystal structure. The structural selection evident in Fig. 6 is directly related to the decrease in the diversity D that we discussed in the previous Section. This type of structural selection represents, generally, the most common form of evidence presented in the literature to support the proposition that the properties of an amorphous material are governed by the stability of a small set of structures. A number of papers [41-43] have reported that the (0,2,8,0) structure in A$_{80}$B$_{20}$ is associated with a relaxation time that is a factor of 3-6 times slower than the average relaxation time of the liquid. An analogous observation has been reported for the (0,0,12,0) structure around Cu in the CuZr mixture by Cheng, Sheng and Ma [44].

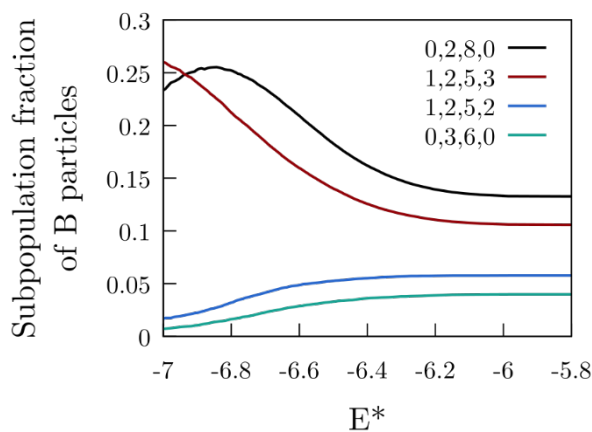

**Figure 6.** Plots of the fraction of B-centred structures in $A_{80}B_{20}$ as a function of the associated subpopulation of particles sampled where the subpopulation is defined by those particles with potential energy below an energy threshold $E^*$.

Are these types of correlations sufficient to claim a causal link between the favoured structures and the physical effect? In Fig. 7 we sketch three different scenarios regarding structure-property relations. The ideal of a perfect causal resolution (Fig. 7a) involves a one-to-one map of each structure to a specific range of property values. This scenario is sufficient to establish that the measured structural property is the cause of the observed property. The opposite of this ideal is that the chosen structural resolution has no correlation with the property (i.e. the null hypothesis) as sketched in Fig.7b. A third possibility, on that better describes correlations of the type actually observed (i.e. in Fig. 6 and refs. [39-42]), is the partial resolution depicted in Fig. 7c. Here an extreme of property values (e.g. lowest mobility, lowest energy, etc) is associated with only one (or a small number) of structures while the structures themselves might contribute to a range of property values. This partial resolution as sketched in Fig. 7c and as demonstrated in Fig. 6 will tend to have low utility (as defined here), despite exhibiting strong structural selectivity at the extreme of the property distribution, due to the significant global overlap of the distribution of property values (as shown in Fig. 5). So what can we conclude from the observation of a scenario like Fig. 7c? We suggest the following: *The structures selected for by the property extreme are a component of the structures responsible for the distribution of the property value but they are not a complete description of the structures responsible.* In the case of the $A_{80}B_{20}$ mixture, for example, the (0,2,8,0) B environment does indeed contribute to stability – both mechanical and kinetic – but only when some other, unmeasured and, presumably, nonlocal, structural conditions apply. The utility we have defined can, in this context, be regarded as a rough measure of the degree to which a given structural classification (like the strictly local one provided by the Voronoi analysis) discards important information.

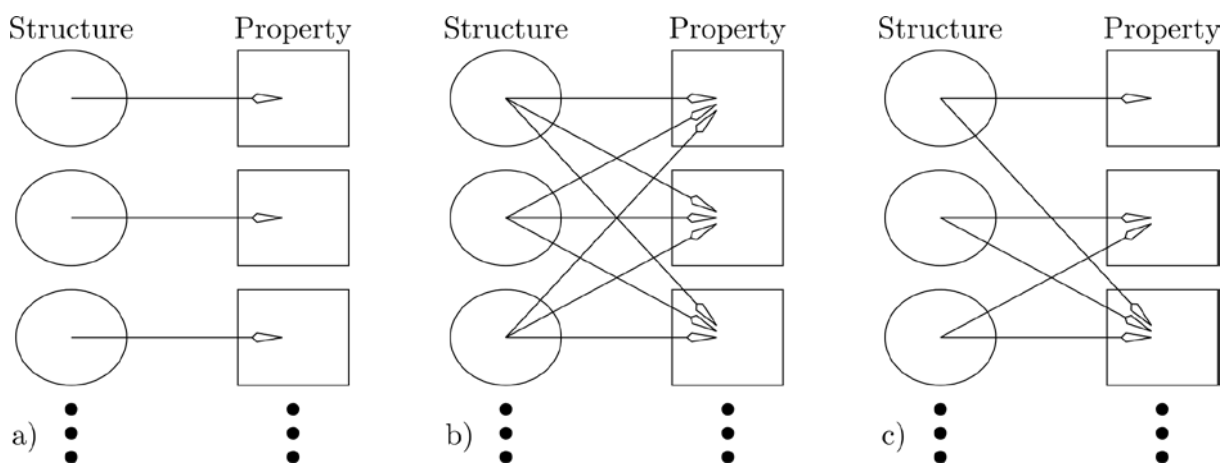

**Figure 7.** Diagramatic representations of possible patterns of causal connection between structure and property. a) Perfect causal resolution (i.e. $U_X = 1$) where each structure gives

rise to a distinct range of property values. b) The null hypothesis (i.e. $U_X = 0$) in which each structure is associated with a broad range of property values. c) The partial resolution ($0 < U_X < 1$) in which extremes of property values are associated with a single structure but that structure itself contributes to a range of property values.

### 3.3 Structure and the Proximity of a Crystalline Phase

The most common use proposed for amorphous structures has been to rationalise the absence of crystallization. Exactly how this rationalisation is to be achieved is not clear. (A good example of what is required to establish a connection between structure and crystallization is provided by Taffs and Royall [45].) To examine the structure of a glass and try to infer what aspect of the structure might have contributed to the non-observation of ordering is flawed as a logical proposition, akin to trying to explain in cards why one didn't get dealt four Kings by examining the distribution of cards that one did receive. The frequency of crystal-like fluctuations would provide a clear basis for explaining the observed crystallization rate. Obtaining statistics of the crystal-like structural fluctuations, however, is problematic because of a) the rare occurrence of these structures in liquids, and b) the large fluctuations we would expect around such high symmetry structures in finite size clusters. These difficulties are compounded by the possible presence of metastable crystalline alternatives (i.e. polymorphs).

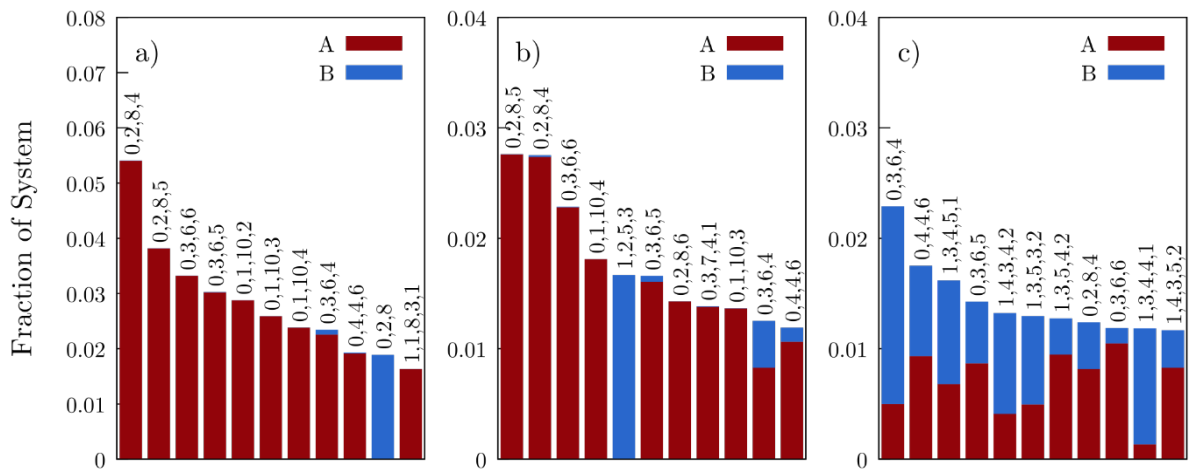

**Figure 8.** Comparison of the distribution of the different species, A and B, among local structures for three different compositions of the KA model: a) $A_{80}B_{20}$, b) $A_{66}B_{33}$ and c) $A_{50}B_{50}$.

Given the difficulties of a direct examination of the statistics of the explicit crystal-like structure in a liquid, it is useful to cast around for more general crystal-related features. In the case of a binary alloy crystallization into an AB crystal, one aspect of the crystal structure that is insensitive to structural details is that in most AB crystal structures, the A and B species occupy identical structural sites. It follows that the degree to which local structures are inhabited by both atomic species in a liquid known to crystallize to an AB structure could be regarded as a measure of susceptibility to nucleate. Note that this condition does not require us to choose any specific structures; the selection is left up to the liquid. In Fig. 8 we present distributions of the most frequent structures in the KA mixture at three different compositions. As we approach the equimolar concentration, we see the two species A and B, increasingly sharing common local structures. Indeed, the equimolar AB liquid freezes into an AB crystal and does so far more rapidly than at the other two concentrations [27].

To measure this degree of shared structures, we shall define a quantity $f_x$ as the average weighted fraction of mutual participation in common structure by the two chemical species in a binary alloy as follows,

$$f_x = \frac{4}{N} \sum_i \frac{n_i^A n_i^B}{n_i^A + n_i^B} \tag{8}$$

where $n_i^A$ = the number of A particles with structure i. If the chemical species separate completely into distinct structures, $f_x = 0$, while, if the two species are equally represented in each structure, the $f_x = 1$ (for an equimolar mixture). Values of $f_x$ for the CuZr and KA mixtures are presented in Table 1. We find that the $A_{50}B_{50}$ KA mixture has value of $f_x = 0.55$ while the large particle rich mixtures have $f_x$'s $< 0.1$. The difference in $f_x$ correlates well with the difference in crystallization kinetics; rapid in the case of the $A_{50}B_{50}$ mixture and slow in the $A_{80}B_{20}$ case. The trend towards shared structures for the two species is prevalent in the $A_{50}B_{50}$ mixture well before any sign of crystallinity, a non-trivial result. In concluding, we reiterate that a) $f_x$ does not require that we make any selection of specific target structures, and b) we have established a correlation between slow crystallization and the absence of a crystal-like structure rather than to the presence of some non-crystalline structure.

## 4. Discussion

To assess the utility of a particular structure classification of an amorphous material requires that we entertain the possibility of the null hypothesis, i.e. that our measure of structure is not useful. If this hypothesis seems to ignore the history of successes of structure-property correlations in science, we stress that these successes generally refer to the use of the total structure as obtained from crystalline materials. In this paper we do not challenge this position but emphasise that, in amorphous materials, the total structure is typically inaccessible and, even when available as in colloid microscopy or simulations, is simply too complex to be of intelligible use. It is the incomplete character of structural description in amorphous materials that raises the question of utility. This point is important. There is a growing body of evidence for the existence of heterogeneous dynamical and material properties of an amorphous structure [46]. Such observations provide clear evidence of *some* sort of structural control without providing any clear indication of which (if any) choice of local structural classification might successfully capture this structural control and the associated structure-property correlation.

The essential character of amorphous structure – the large multiplicity of local arrangements – has been presented here as an explicit structural measure in the form of the diversity D.

Evaluating D for glass forming alloys, we find D > 100, in clear contrast to crystalline states which are roughly limited to D < 10. The quantity D provides an explicit measure of the significance of any particular local structure. The observed values of D raises a number of questions. What manner of materials occupy the intermediate structural diversities i.e. 10 < D <100? Should a large diversity be regarded as a fundamental feature of a material or just the signature of a poor choice of structural classification? The latter question can be answered, in part, by testing a range of structural measures so that if D remains relatively constant then one can assume that the value of D does indeed reflect some intrinsic feature of the structure. How does the diversity D change as we consider the subpopulations characterised by some restricted set of property values? In the paper have considered this question in the context of constraint and found a substantial reduction of diversity as we consider only structures with increasing degree of particle localization. Our analysis demonstrated that this loss of diversity could be quantitatively attributed to both species selectivity and structural selectivity. The structural selectivity takes the form of particular Voronoi polyhedra dominating the structure of the most constrained subpopulation – a result qualitatively similar to previous reports of structural significance. Using our utility index, we can qualify this observation by noting that the structural measure can have a low utility even while exhibiting this high selectivity for the extreme of a property. This is a signature of an *incomplete* structural descriptor. What this means is, for example, while the structure (0,2,8,0) corresponds to a significant fraction of B particles in the KA mixture with low energy and low mobility, the identification of a given particle as being (0,2,8,0) tells us little about its stability relative to particles with other structures. Expanding the structural measure is a non-trivial task. Machine learning [47] has been proposed as a strategy for addressing this incompleteness. In this approach, the weighting of different types of structural data is adjusted to maximise coincidence with some selected property, e.g. local dynamics. An unresolved issue with such combinatoric approaches is to identify exactly what use (in the sense we discuss in Section 1) the resulting structure serves.

Based on the analysis presented here, we conclude that the Voronoi analysis is of limited utility in the description of the two alloys selected for this study. We emphasise that this conclusion refers specifically to our choice of Voronoi polyhedral and our glass forming liquids. That said, we would expect that, in liquids with a large diversity, no local measure of topology or geometry is likely to fare much better. In such situations, where do we go with structural analysis? A useful approach is to consider exactly what are the basic consequences of structure. Previously [48] it has been argued that constraint (i.e. particle localization), rather than the structures responsible for the localization, is sufficient to account for the rigidity of a material, ordered or amorphous. While an amorphous material may have a high diversity of local topologies, the variety of constraints experienced by particles might be much less diverse. Structural measures based on incomplete coordination shells [49,50] are one choice of local structure aimed to connect constraint and structure. An example of a measure of constraint that forgoes any explicit treatment of structure is the point-to-set algorithm [51] where a group of particles are constrained. The measured influence of this constraint on unconstrained particles has been referred to as an 'agnostic' structure – 'agnostic' in the sense that no specific choice of structural classification was required. The potential pitfall of defining structure by its consequence is the circular argument. The point-to-set structure, for example, cannot be used to explain the constraints that were used to define it in the first place.

**5. Conclusion**

In conclusion, we have presented three new measures of structure in amorphous materials that do not rely on intuition or prior assumption regarding the significance of special local arrangements. In treating all aspects of an amorphous structure equally, dependent only on the frequency with which they occur, we have sought to demonstrate how a new class of questions can be put to materials with the goal of moving beyond the demonstration of correlation between structure and property in amorphous materials and establishing those aspects of amorphous structure that provide quantifiable benefit in rationalising the underlying causes of observed material behaviour.

**Acknowledgements**

L.H.D and Y.J.W are supported by the NSFC (Nos. 11672299, 11522221, and 11790292), the National Key Research and Development Program of China (Nos. 2017YFB0702003 and 2017YFB0701502), the Strategic Priority Research Program (No. XDB22040303), the Key Research Program of Frontier Sciences (No. QYZDJSSW-JSC011), and the Youth Innovation Promotion Association of CAS. J.D. thanks the VILLUM Foundation's Matter project (16515) for support. I.D. and P.H. acknowledge the support of the University of Sydney and Chemistry High Performance Computing Facilities.